\documentstyle[12pt]{article}
\def \sect #1 {\setcounter{equation} 0\section{#1}}
\def \be  {\begin{equation}}
\def \ee  {\end{equation}}
\def \ba  {\begin{eqnarray}}
\def \ea  {\end{eqnarray}}
\def \baa {\begin{eqnarray*}}
\def \eaa {\end{eqnarray*}}
\def \bb  {\begin{thebibliography}}
\def \eb  {\end{thebibliography}}
\newcommand \ci [1] {\cite{#1}}
\newcommand \bi [1] {\bibitem{#1}}
\def \lab #1 {\label{#1}}
\newcommand\re[1]{(\ref{#1})}

\newcommand\lr[1]{{\left({#1}\right)}}

\newcommand \VEV [1] {\left\langle{#1}\right\rangle}

\def \e {\mbox{e}}

\newcommand{\as}{\ifmmode\alpha_{\rm s}\else{$\alpha_{\rm s}$}\fi}

\def \DY {_{\rm DY}}
\begin{document}

\def\thefootnote{\fnsymbol{footnote}}
\thispagestyle{empty}
\hfill\parbox{35mm}{{\sc ITP--SB--95--17}\par
                        hep-ph/9505391  \par
                        May, 1995}

\vspace*{40mm}
\begin{center}
{\LARGE Universality of infrared renormalons
\\[3mm]
in hadronic cross sections}
\par\vspace*{15mm}\par
{\large Gregory~P.~Korchemsky}%
\footnote{On leave from the Laboratory of
Theoretical Physics, JINR, Dubna, Russia}
and {\large George~Sterman}
\par\bigskip\par\medskip

{\em Institute for Theoretical Physics, \par
State University of New York at Stony Brook, \par
Stony Brook, New York 11794 -- 3840, U.S.A.}
\end{center}
\vspace*{15mm}

\begin{abstract}
We discuss the role of infrared renormalons in power corrections to hadronic
cross sections, including their universality. We show how perturbative
renormalon structure arises near kinematic boundaries in the thrust
distribution, the Drell-Yan cross section and radiative B decays. The leading
infrared renormalon in each case is associated with jet evolution. Demanding
that the combination of perturbative and nonperturbative contributions be
well-defined, we infer the form of the leading power corrections to each cross
section. This is at the level $1/Q$ for the thrust distribution and Drell-Yan
process, and $1/m^2$ for B decay. We discuss the universality of $1/Q$
corrections between Drell-Yan and thrust, and conclude that it is approximate,
due to contributions from multijet configurations that may differ in the two
cases.
\end{abstract}

\vfill

\centerline{\it to appear in the Proceedings of the XXXth Rencontres
de Moriond}
\centerline{\it ``QCD and High Energy Hadronic Interactions'',
               Les Arcs, France, March 1995}

\newpage
\def\thefootnote{\arabic{footnote}}
\setcounter{footnote} 0

\newpage

\noindent {\bf 1. Introduction.}
Our current understanding of hard processes in QCD is based on
 factorization
(the ``improved the parton model"), which allows us to calculate cross
sections in the high energy limit $Q^2\to \infty$ (with $Q^2$ a large
kinematic scale) as a convolution of nonperturbative parton distribution
functions and short distance partonic subprocesses.  The latter are
calculable in perturbative QCD as a series in $\as(Q^2)$. Although this
procedure is adequate for a variety of processes, there is a wealth
of data for which low-order perturbation theory alone is not enough, most
notably the transverse momentum distributions of $W^\pm$ and $Z^0$ in the
Drell--Yan process \ci{DYpt,KT,AEM,dy} and direct photon and diphoton
production \ci{R,o}, as well as event shapes in $\e^+\e^-$--annihilation
\ci{cttw,w1}. As we approach the edge of the phase space in these processes,
we observe a disagreement with next-to-leading perturbative QCD calculations
that can be attributed to nonperturbative (or hadronization) effects, our
understanding
of which is still incomplete. Remarkably enough, in some cases
one may achieve a better agreement with experimental data by including
additional factors into parton model predictions ``by hand''. These factors
take into account nonperturbative effects in the form of power
corrections in $1/Q$. In this paper we would like to describe a simple method
\ci{KS} which allows us to understand the appearence of additional
nonperturbative factors in resummed cross sections. It is based on the
analysis of ambiguities of the perturbative series in QCD associated with the
so--called infrared renormalons \ci{IR1,IR2}.

In the special case of processes (like the total cross section for
$\e^+\e^-\to\rm hadrons$, deep-inelastic scattering, etc.), which admit the
operator product expansion (OPE), nonperturbative power (``high twist'')
corrections to a physical distribution $\sigma(Q)$ can be parameterized by
dimensional nonperturbative scales $\sigma_n$ as
\be
\sigma(Q) = \sigma_{\rm pert}(Q) + \sum_{n=n_0}^\infty \frac{\sigma_n}{Q^{n}}
\,.
\lab{n0}
\ee
Here the first term corresponds to the perturbative contribution
$\sigma_{\rm pert}=\sum_k c_k \lr{\as(Q^2)}^k$ and the leading power
corrections
appear as $\sim Q^{-n_0}$. While the complete sum in \re{n0} defines a physical
quantity $\sigma(Q)$, individual terms are not well defined, either due to
ambiguities in the definition of nonperturbative quantities ${\sigma_n}\sim
\Lambda_{\rm QCD}^n$ or due to divergences of $\sigma_{\rm pert}$ related
to the factorial growth of the coefficients $c_k \sim \beta_0^k k!$ (with
$\beta_0=11-2/3 n_f$) at higher orders of perturbative QCD. As a result, to
give a meaning to the perturbative (as well as nonperturbative) contribution
to $\sigma(Q)$ one has to specify the way in which we ``regularize''
divergences of $\sigma_{\rm pert}$. Different prescriptions lead to
different
results for $\sigma_{\rm pert}$ which
eliminate ambiguities at the level of power
corrections $\Lambda_{\rm QCD}^n/Q^n$. It is important for us that ambiguities
of the perturbative contribution $\sigma_{\rm pert}$ are compensated in \re{n0}
by ambiguities in the definition of nonperturbative scales $\sigma_n$. Using
this property we can explore the ambiguities of $\sigma_{\rm pert}$ to
identify the structure of nonperturbative corrections, and in particular the
level $n_0$ at which the leading power corrections appear in \re{n0}.
A justification for this scheme is
partly in our expectation that there is a unique theory of
QCD involving perturbative and nonperturbative effects, and also
that for processes in which the OPE is
applicable and $n_0$ is known explicitly, the analysis of perturbative
ambiguities leads to the same result \ci{IR2}.

Let us apply the same approach to hadronic processes to which the OPE is not
applicable. To be concrete, we will consider the following three different
processes:  the thrust distribution $d\sigma/d T$ in $\e^+\e^-$ annihilation,
the energy distribution of lepton pairs in the Drell-Yan process and
the differential rate of the radiative decay of the B meson. One of the reasons
why the OPE does not apply
directly to these processes is that they involve two scales
$Q^2$ and $q^2$ with $Q^2 \gg q^2 \gg \Lambda_{\rm QCD}$.
(We shall see below, however, an important role for
the OPE in heavy quark effective theory \ci{HQETOPE}.)
As a result, the
short--distance partonic subprocess includes large double--logarithmic
(Sudakov) corrections $(\as(Q^2)\ln^2(Q^2/q^2))^n$, due to the enhancement
of soft gluons with momenta $k\sim q$. Our strategy will be the following
\ci{KS}. We first apply factorization, and perform
resummation of large perturbative logarithmic corrections. The resulting
resummed expressions involve integrals of the running constant at very low
scales, given by soft gluon transverse momenta. This makes the perturbative
series sensitive to the way in which we treat singularities of the coupling
constant. Once we identify the ambiguities of the Sudakov resummed
perturbative expansions, we add nonperturbative corrections in order
to compensate them and make the final expressions well-defined.
Although the analysis of perturbation theory cannot give us the absolute
normalization of nonperturbative corrections, we may use it to parameterize
nonperturbative effects by introducing new scales with dimensions fixed
by the positions of infrared renormalons.

\bigskip\noindent
{\bf 2. Sudakov resummation.}
For $\e^+\e^-$ annihilation, the thrust is defined in terms of
the momenta $k_j$ of final state particles as
\be
T={\rm max}_{_{\mbox{$\,\vec n$}}}\
 \frac{\sum_j |\vec k_j\cdot \vec n|}{\sum_j |\vec k_j|}
\,.
\ee
It takes a maximum value $T=1$ for two infinitely narrow jets. For $1-T$
small the final state consists of two jets, with invariant masses $k^2$ and
$\bar k^2$, respectively, and with energy $Q$ in the center-of-mass
frame, $Q^2\gg k^2,\; {\bar k^2}$. Then, in the $T\to 1$ limit the thrust
is given by \ci{cttw}
\be
T=1 - \frac{k^2}{Q^2} - \frac{\bar k^2}{Q^2}\, ,
\ee
and for the thrust distribution one can
derive the factorized form \ci{cttw}
\be
\frac1{\sigma_{\rm tot}} \frac{d\sigma}{dT} \stackrel{T\sim 1}{=}
\int_0^\infty dk^2 \int_0^\infty d\bar k^2 \ J_{\rm q}(Q^2,k^2)
J_{\rm \bar q}(Q^2,\bar k^2) \
\delta\lr{1-T-\frac{k^2}{Q^2} - \frac{\bar k^2}{Q^2}}\, ,
\lab{Tdis}
\ee
where $\sigma_{\rm tot}$ is the total cross section of $\e^+\e^-\to
\rm hadrons$. Here, the jet distribution $J_{\rm q}(Q^2,k^2)$ describes
the probability for a quark with energy $Q$ and invariant mass $k^2\ll Q^2$
to create a jet of collinear particles accompanied by soft gluon radiation
\ci{jet1,jet2}. In the Born approximation, $J_{\rm q}^{(0)}=\delta(k^2)$,
while the emission of soft and collinear particles by the ``parent'' quark
leads to
large logarithmic corrections $\as^n/k^2 \ln^{2n-1} Q^2/k^2$ to
$J_{\rm q}$. They
can be resummed using evolution equation techniques, and the final expression
for $J_{\rm q}$ can be represented to next-to-leading logarithmic
accuracy in the form of a Laplace transform \ci{jet1,jet2,cttw},
\be
\int_0^\infty dk^2 \e^{-\nu k^2/Q^2} J_{\rm q}(Q^2,k^2)
=\exp\left\{-\int_0^1\frac{du}{u}\lr{1-\e^{-u\nu}}
\left[
\int_{u^2Q^2}^{uQ^2}\frac{d k_t^2}{k_t^2} \Gamma_{\rm cusp}(\as(k_t^2))
+\gamma(\as(uQ^2))
\right]
\right\}\,.
\lab{jet}
\ee
An identical expression applies to the antiquark jet distribution. Here,
$\nu$ is the transform parameter, and $\Gamma_{\rm cusp}(\as)$ and
$\gamma(\as)$ are anomalous dimensions, known to the lowest orders of
perturbative QCD. We notice that the jet distribution \re{jet}
satisfies the normalization condition
\be
\int_0^\infty dk^2  J_{\rm q}(Q^2,k^2) = 1\, ,
\ee
which guarantees that $\int dT \frac{d\sigma}{dT} = \sigma_{\rm tot}$.
This means, that in agreement with the OPE for $\sigma_{\rm tot}(Q)$,
large perturbative (as well as nonperturbative) corrections to the
differential cross-section $d\sigma/dT$ described by \re{Tdis} and \re{jet}
cancel in the total cross-section. Using \re{Tdis} we find the following
expression for the thrust distribution
\be
\VEV{\e^{-\nu(1-T)}}\equiv
\int_0^1 d T \frac1{\sigma}\frac{d \sigma}{d T} \e^{-\nu(1-T)}
=\left[
\int_0^\infty dk^2 \e^{-\nu k^2/Q^2} J_{\rm q}(Q^2,k^2)
\right]^2\, .
\lab{nu}
\ee
Together with \re{jet} this relation resums all large perturbative
corrections which describe the effect of Sudakov suppression of the
end--point region $T\sim 1$ in the thrust distribution $d\sigma/dT$.
We notice, however, that in \re{jet} the integration over the transverse
momenta of soft gluons is potentially divergent due to singularities of the
coupling constant $\as(k_t^2)$, and it is this property of the Sudakov
resummed perturbative expression which we are going to use to identify
nonperturbative effects.

As a second example, we consider the distribution of the lepton pairs
in the Drell-Yan process $h_1 + h_2 \to \ell^+\ell^- +X$ with respect
to their invariant mass $Q^2$ at the edge of phase space as $Q^2$
approaches the invariant energy of the incoming hadrons $\sqrt{s}$.
As $\tau=Q^2/s \to 1$,
the final state in the partonic subprocess is dominated by soft gluons with
total energy $\sim (1-\tau)Q$, which lead to the appearence of large
Sudakov perturbative corrections to the short-distance partonic cross-section
\ci{jet1,jet2,KM}. The proper quantities to study in this case are the moments
of the differential cross-section $d\sigma_{_{\rm DY}}/d\tau$, normalized to
the moments of a structure function of deep-inelastic scattering $F(x)$,
\be
\Delta_n=\sigma_n/F_n^2\,,\qquad
\sigma_n=\int_0^1 d\tau \tau^n \frac{d\sigma_{\DY}}{d\tau}\,,\quad
F_n=\int_0^1 dx x^n F(x)\,.
\ee
Each of these quantities may be factorized into the product of moments of
a parton distribution function and moments of a short-distance partonic
cross-section. The parton distributions cancel in the ratio, $\Delta_n$,
which makes it possible to evaluate $\Delta_n$ perturbatively. For
large $n$, the ratio $\Delta_n$ has large Sudakov perturbative
corrections $(\as\ln^2 n)^k$, whose origin can be traced back to the
asymptotics of $d\sigma_{\DY}/d\tau$ and $F(x)$ as $\tau \to 1$ and $x\to 1$,
respectively. In the Drell-Yan process, large corrections to
$d\sigma_{\DY}/d\tau$ are caused by soft
gluons emitted from the initial state by the quark and antiquark before they
annihilate. At the same time, in deep-inelastic scattering large
corrections to $F(x)$ as $x\to 1$ arise from soft gluon emission from the
incoming quark, as
well as from the decay of the scattered quark with energy $\sim Q$
and small invariant mass $k^2=Q^2(1-x)/x$ into a jet of collinear and
soft particles in the final state. We recognize that the latter subprocess
is described
by the same jet distribution $J_{\rm q}(Q^2,k^2)$ which entered into
the analysis of the thrust distribution. Moreover, in the ratio of the moments
$\Delta_n$ the contributions
of soft gluons emitted by quarks in the initial state in the Drell-Yan
process and deep-inelastic scattering
cancel, and the only contribution which
survives is that of the jet distribution associated with the outgoing quark in
deep-inelastic scattering. As a result,
for large $n$, $\Delta_n(Q)$ is given by
\ba
\Delta_n(Q) & = & H(\as(Q^2))
     \left[Q^2 \int_0^1 dx x^n J_{\rm q}(Q^2,Q^2(1-x)/x) \right]^{-2}
\nonumber
\\
            &=& H(\as(Q^2))
     \left[\int_0^\infty dk^2 \e^{-n k^2/Q^2} J_{\rm q}(Q^2,k^2) \right]^{-2}
+{\cal O}(1/n)\, ,
\lab{delta}
\ea
where $H = 1 + \sum_k h_k\ \as^k(Q^2)$ comes from the
contributions of hard virtual
gluons with momenta $k\sim Q$, and where the jet distribution has been defined
in \re{jet}. Comparing this expression with \re{nu} we notice that the
resummed expressions for different physical quantities, $\Delta_n(Q)$ and
$\VEV{\e^{-n(1-T)}}$, have the same form.

As a last example we consider the differential rate
for the inclusive radiative
decay ${\rm B}\to \gamma X_s$ in the end-point region of the photon spectrum,
in the
limit when the mass of the s quark is neglected. In the rest frame of the
B meson, we define a scaling variable $x$ as the ratio of the photon energy
to the mass of the b quark, $x=2E_\gamma/m$.
That is, we are interested in the
inclusive distribution $d\Gamma/dx$ with $x\sim 1$. In the heavy quark limit,
$m\to\infty$, we may
factorize the decay distribution into a convolution
of the b quark distribution function in the B meson and a short distance
partonic subprocess ${\rm b}\to {\rm s}\gamma$. The latter gets large
perturbative
corrections $\sim [\as^n/(1-x)]\; \ln^{2n-1}(1-x)$ in the end-point region,
 which originate from the propagation of the s quark into the final
state with a large energy $\sim m/2$ and a small invariant mass,
$k^2=m^2(1-x)$. The s quark creates a jet of collinear particles accompanied
by soft radiation, which is described by the same jet distribution
$J_{\rm q}(m^2,m^2(1-x))$ that we encountered above. Introducing the moments
of the differential rate $d\Gamma/dx$ as
\be
{\cal M}_n({\rm B}\to\gamma X_s)=\int_0^{x_{\rm max}} dx\ x^n \frac 1{\Gamma}
\frac{d\Gamma}{dx}\, ,
\lab{x}
\ee
and performing the Sudakov resummation of large perturbative corrections
to ${\cal M}_n$, for large moments $n$ we may represent the result in
the following factorized form \ci{KS1}
\be
{\cal M}_n= f_n^{(0)}\ H(\as(m^2))\ J_n\ S_n + {\cal O}(1/n)\, ,
\lab{decay}
\ee
where $f_n^{(0)}$ is a nonperturbative parameter describing the distribution
of the b quark in the B meson, and $H$ comes from hard gluon corrections to
the effective vertex ${\rm b}\to{\rm s}\gamma$.  It is similar to the
analogous factor in \re{delta}. $J_n$ takes into account all effects of the
s quark fragmentation into a jet and is equal to
\be
J_n = m^2 \int_0^1 dx x^n J_{\rm q}(m^2,m^2(1-x))
   = \int_0^\infty dk^2 \e^{-n k^2/m^2} J_{\rm q}(m^2,k^2) + {\cal O}(1/n)\, ,
\ee
with the jet distribution given by \re{jet}. There is an additional factor
$S_n$ in \re{decay} which takes into account the nonleading contribution of the
soft gluons emitted by the incoming b quark. It is defined as
\be
S_n = \exp\left\{-\int_0^1\frac{du}{u}\lr{1-\e^{-nu}} \Gamma(\as(u^2m^2))
\right\}\, ,
\lab{sn}
\ee
with $\Gamma(\as)$ an anomalous dimension \ci{KS1}. Comparing \re{sn} and
\re{jet}, we notice that the contribution of $S_n$ to the moments
${\cal M}_n$ is subleading
with respect to that of jet subprocess, $J_n$,
so that it can be reabsorbed
into a redefinition of the anomalous dimensions $\Gamma_{\rm cusp}$
and $\gamma$ to higher orders of perturbation theory.

Thus, for three processes considered above the Sudakov resummation of
large perturbative corrections gives similar expressions, \re{nu},
\re{delta} and \re{decay}. The reason for this universality is that in
all cases large perturbative corrections
came from the same jet subprocess
which can be described by the distribution
function $J_{\rm q}$ defined in \re{jet}.

\bigskip
\noindent
{\bf 3. Power corrections from infrared renormalons.}
Expressions \re{nu}, \re{delta} and \re{decay} were found after resummation
of large Sudakov logarithmic corrections to all orders of perturbation theory.
This does not guarantee, however, that the resummed perturbative series
are convergent. Indeed, it is clear from \re{jet} that the integral over
the transverse momenta of soft gluons is divergent at small values of $k_t^2$.
In the perturbative expansion of the exponent in the r.h.s.\ of \re{jet}
in powers of $\as(Q^2)$ these
divergences manifest themselves in factorial growth of the coefficients
$\sim \beta_0^k k!$. This tells us that, taken alone, resummed perturbative
predictions \re{nu}, \re{delta} and \re{decay}
are not well defined. Since this problem arises from small momenta
region of soft gluons, we do not expect perturbative QCD alone to suffice.
Rather, perturbative expressions should be supplemented by nonperturbative
corrections, and it is their sum that should give unique predictions for
physical
quantities. Although we do not know in general how to evaluate nonperturbative
effects from first principles,
we may explore the ambiguities of perturbative expressions like \re{jet}
to indentify potential sources of nonperturbative corrections.

Let us identify the leading infrared renormalon singularity in the expression
for the jet distribution \re{jet}. It is defined by the first term in the
exponent, in which we interchange the $k_t$ and $u$ integrals and expand
the result for small transverse momentum to get \ci{KS}
\be
\exp\left\{-
\frac{2\nu}{Q} \int_0^{Q} d k_t \ \Gamma_{\rm cusp}(\as(k_t^2))
\left[{1+{\cal O}(k_t/Q)}\right]
\right\}\, .
\lab{form}
\ee
We conclude that the resummed perturbative expression for the jet distribution
\re{jet} has its leading infrared renormalon singularity at the level of
a $1/Q$ power correction,
\be
\int_0^\infty d k^2 \ \e^{-\nu k^2/Q^2} J_{\rm q}(Q^2,k^2)\bigg|_{\rm nonpert}
\sim
\exp\lr{-\nu A /Q+{\cal O}(1/Q^2)}\, ,
\lab{Q}
\ee
where the dimensional parameter $A$ includes both perturbative and
nonperturbative contributions. Although the latter may depend on the
process in which the jet distribution enters, the former have a universal
form
$\lr{\sim 2\int_0^{Q} d k_t \ \Gamma_{\rm cusp}(\as(k_t^2)}$,
which follows from \re{form}. Applying \re{Q} to \re{nu} and \re{delta}
we estimate the magnitudes of nonperturbative effects in the thrust
distribution and in the Drell--Yan process as
\be
\VEV{\e^{-\nu (1-T)}}\bigg|_{\rm nonpert} \sim
\e^{-2\nu A_T/Q}
\,,\qquad
\Delta_n(Q)\bigg|_{\rm nonpert} \sim
\e^{2nA_{\DY}/Q}
\,.
\ee
This analysis of perturbative ambiguities,
and their relation to nonlocal operators, suggests \ci{KS}
that the contribution of the leading infrared renormalon to the parameters
$A_T$ and $A_{\DY}$ is the same\footnote{This result has been
recently confirmed in \ci{T,AZ}},
although it does not allow us to draw definitive conclusions on
the magnitudes of these parameters.

One may try to apply the same result to the moments of the B meson
differential rate, eq.\re{decay}. There is, however, an important
difference from the previous cases. For B decay, the energy scale $Q$, which
is a kinematic invariant for the cross sections (center of mass energy for
thrust and lepton mass in the Drell--Yan process), should be replaced by the b
quark mass $m$. The $1/m$ contribution is given by an operator expectation
value in heavy quark effective theory that vanishes by the heavy quark's
equation of motion.
This result may also be seen directly in perturbation theory. It turns out
\ci{mass} that the definition of $m$ is also ambiguous due to the presence
of an ultraviolet renormalon in the perturbative series for $m$. Therefore,
analyzing perturbative ambiguities to the moments ${\cal M}_n$ of
the differential rate, we have to take into account contributions of
infrared renormalons to the jet distribution \re{Q} and that of the
ultraviolet renormalon to the b quark mass, $m \to m+\delta m$. The net effect
of the latter ambiguity can be reduced to a renormalization of the scaling
variable $x \to x (1-\delta m/m)$ in \re{x} and the renormalon
contributions to the moment ${\cal M}_n$ can be represented in the large
$m$ limit as
\be
{\cal M}_n\bigg|_{\rm nonpert} \sim \exp(-nA /m - n\delta m/m)
\ee
with parameter $A$ defined in \re{Q}. The
absence of $1/m$ corrections implies \ci{GK} that their sum vanishes,
$A+\delta m=0$, and
perturbative ambiguities will first appear in ${\cal M}_n$ at the level
${\cal O}(1/m^2)$ power corrections.

We stress that although the resummed perturbative expressions for the thrust
distribution and in the B meson decay, eqs. \re{nu} and \re{decay}, look very
similar after identification of the energy $Q$ with heavy quark mass $m$,
their properties with respect to renormalon ambiguities are completely
different, due to ambiguities in the definition of heavy quark mass $m$.
However, keeping in mind the situation with B decay, we may interpret
nonperturbative $1/Q$ corrections to the thrust distribution as
a renormalization of
the scaling variable $1-T$, or equivalently the invariant masses $k^2$ and
$\bar k^2$ of the jets. This implies, in particular,
that the emission of nonperturbative soft gluons by
the quark and antiquark makes the jets
in the final state wider, $(k^2+\bar k^2)/Q^2 >A_T/Q$, and the phase space
for the thrust becomes smaller, $T < 1-A_T/Q$. Indeed, using the expression
\re{nu} for the thrust distribution, and expanding it in powers of $\nu$, one
can estimate the nonperturbative correction to the average thrust from
two--jet configurations,
\be
\VEV{1-T}\bigg|_{\rm nonpert} \stackrel{{\rm 2-jet}}{\sim}
A_T/Q\,.
\lab{1-T}
\ee
In contrast to the thrust distribution \re{nu} at $\nu \gg 1$, the average
thrust $\VEV{1-T}$ gets leading contributions not only from 2--jet events
but from multi--jet events as well. The latter should be analyzed
separately, and there is no reason to expect that their contribution
to the average thrust will be subleading
by a power with respect to that in \re{1-T}.
Indeed, an analysis of renormalon ambiguities to perturbative series
corresponding to multi--jet final state shows \ci{in}
that leading nonperturbative
corrections to $\VEV{1-T}$ appear at the same level $1/Q$ as in the
2--jet events, although suppressed in general by powers of $\alpha_s(Q)$.
Thus, the total leading nonperturbative correction to the
average thrust has the same form \re{1-T} but with the scale $A_T$ modified
by the contributions of multi--jet events. Contributions
of infrared renormalons in $1/Q-$power corrections to $\VEV{1-T}$
associated with multi--jet events prevent an exact equality between
perturbative ambiguities in the average thrust and in the Drell--Yan
process.

\bigskip
\noindent
{\bf 4. Conclusions and perspective.}
The analysis of ambiguities associated with infrared renormalons allows us
to gain insight into the structure of nonperturbative corrections in
hadronic processes to which the OPE is not applicable. Performing Sudakov
resummation of large perturbative corrections,
 we found that in all three cases, the thrust distribution, the Drell--Yan
process and the differential rate of the B meson decay,
the contribution of
the leading infrared renormalon has a universal form, which suggests that
nonperturbative effects may appear at the level of $1/Q$ (or $1/m$ for the
B meson decay) power corrections. Moreover, the $1/Q-$power corrections
to the thrust distribution and to the Drell--Yan process can be
parameterized by introducing two new nonperturbative scales $A_T$ and
$A_{\DY}$,
which in an analogy with the local condensates in the QCD sum rules can be
related to vacuum expectation value of nonlocal Wilson line operators
\ci{KS,in}.
An important difference occurs for B meson
decay, where the OPE in heavy quark effective
theory ensures that infrared and ultraviolet renormalons to the
$1/m$ power corrections cancel each other, indicating that leading
nonperturbative effects to the differential rate occur at the level of
$1/m^2-$power corrections.

The above considerations can be easily generalized to other processes
which require resummation of large perturbative corrections.
Returning to the example mentioned at the very beginning,
we consider the transverse momentum distribution \ci{KS}
of lepton pairs in the Drell--Yan process $d^2\sigma/d q_t^2 d Q^2$.
In contrast with the previous case, we do not require $Q^2 \sim s$ but
approach the edge of the phase space $q_t^2 \ll Q^2$. In this kinematic region,
the small transverse momentum of the lepton pair is compensated by soft
gluons emitted by the quark and antiquark in
the partonic subprocess. These soft gluons
give rise to large perturbative corrections $\as^k\ln^{2k}(Q^2/q_t^2)$,
which can be resummed to all orders of perturbation theory into a Sudakov
form factor \ci{DYpt}. As above, the
resulting
resummed expressions for the transerse momentum distribution suffers from
infrared renormalon ambiguities which give rise to nonperturbative corrections.
The leading nonperturbative corrections have a simple physical meaning \ci{KS}.
They introduce an additional broading into the transverse momentum
$k_t-$distribution of incoming quarks with a gaussian weight.  In
this fashion, we arrive at the form originally proposed by Collins
and Soper for nonperturbative contributions to the Drell-Yan
$q_t$--distribution \ci{DYpt},
\be
\frac1{4\pi\sigma^2}
\exp\lr{-\frac{k_t^2}{4\sigma^2}}
\,,\qquad \sigma^2= g_1 + g_2 \ln\frac{Q}{2Q_0}\,.
\lab{gauss}
\ee
Here, the width $\sigma$ depends on the invariant energy of the quark and
antiquark in the partonic subprocess, $\hat s =x_1 x_2 s= Q^2$, and the
numerical values of new nonperturbative scales $(g_1, g_2, Q_0)$ have
been estimated \ci{dy} from comparison with experiment as $g_1=0.15\
({\rm GeV})^2,$
$g_2=0.40\ ({\rm GeV})^2$ and $Q_0=2\ {\rm GeV}$.
As we have seen, the infrared renormalons contributing to $1/Q$--power
corrections have universal structure in different processes. The same is
true for the infrared renormalon arising in the Sudakov resummation for the
$q_t-$distribution. One can also show, for example, that resummation of soft
gluon effects for direct photon production at small $x_F=2q_t/\sqrt{s}$
and for diphoton production at small total $q_t$ of the photons, leads to
expressions in which the contribution of the leading infrared renormalon
gives rise to nonperturbative corrections similar to those in \re{gauss}.

One of us (G.P.K.)\ would like to thank the Organizers of the Moriond
Conference for their generous hospitality and support.
This work was supported in part by the National Science Foundation under
grant PHY9309888.

\bb{99}
\bi{DYpt}
     J.C. Collins and D.E. Soper, Nucl. Phys. B193 (1981) 381; {\it ibid \/}
     B197 (1982) 446.
\\   J.C. Collins, D.E. Soper and G. Sterman, Nucl. Phys. B250 (1985), 199.
\bi{KT}
     J. Kodaira and L. Trentadue, Phys. Lett. 112B (1982) 66;
     {\it ibid \/} 123B (1983) 1983.
\bi{AEM}
     G. Altarelli, R.K. Ellis, M. Greco and G. Martinelli,
     Nucl. Phys. B246 (1984) 12.
\bi{dy}
     C.T.H. Davies and W.J. Stirling, Nucl. Phys. B244 (1984) 337;
\\   C.T.H. Davies, B.R. Webber and W.J. Stirling, Nucl. Phys. B256
     (1985) 413;
\\   P.B. Arnold and R.P. Kauffman, Nucl. Phys. B349 (1991) 381.
\bi{R}
     J. Huston et al., MSU-HEP-41027, Jan 1995,  hep-ph/9501230;
\\   A.D. Martin, W.J. Stirling and  R.G. Roberts,   RAL-95-021, Feb 1995,
     hep-ph/9502336.
\bi{o}
     P. Aurenche et al., Nucl. Phys. B399 (1993) 34;
\\   P. Chiappetta, R. Fergani and J.Ph. Guillet, CPT--94/P.3076,
     Dec. 1994.
\bi{cttw}
     S. Catani, L. Trentadue, G. Turnock and B. Webber,
     Nucl. Phys. B407 (1993) 3.
\bi{w1}
     B.R. Webber, Phys. Lett. B339 (1994) 148;
     Cavendish-HEP-94-17, Nov 1994, hep-ph/9411384.
\bi{KS}
     G.P. Korchemsky and G. Sterman,  Nucl. Phys. B437 (1995) 415.
\bi{IR1}
     G. 't Hooft, {\it in\/} The Whys Of Subnuclear Physics, Erice
     1977, ed. A. Zichichi (Plenum, New York, 1977), p. 943;
\\   B. Lautrup, Phys. Lett. 69B (1977) 109;
\\   G. Parisi, Phys. Lett. 76B (1978) 65; Nucl. Phys. B150 (1979) 163;
\\   F. David, Nucl.Phys. B234 (1984) 237; {\it ibid \/}  B263 (1986) 637.
\bi{IR2}
     A.H. Mueller, Nucl. Phys. B250 (1985) 327; {\it in\/} QCD 20 years
     later, Aachen, 1992, ed. P.M.Zerwas and H.A.Kastrup,
     World Scientific, Singapore, 1993, v.1, p.162;
\\   V.I. Zakharov, Nucl. Phys. B385 (1992) 452;
\\   M. Beneke and V.I. Zakharov, Phys. Rev. Lett. 69 (1992) 2472.
\bi{HQETOPE} M. Shifman, Proceedings of the Workshop
    {\it Continuous Advances in QCD}, ed. A. Smilga
    World Scientific, Singapore, 1994,  hep-ph/9405246.
\bi{jet1}
     G. Sterman, Nucl. Phys.  B281 (1987) 310.
\bi{jet2}
     S. Catani and L. Trentadue, Nucl. Phys. B327 (1989) 323; {\it ibid\/}
     B353 (1991) 183.
\bi{KM}
     G.P. Korchemsky and G. Marchesini, Phys. Lett. B313 (1993) 433.
\bi{KS1}
     G.P. Korchemsky and G. Sterman, Phys. Lett. B340 (1994) 96.
\bi{T}
     Yu.L. Dokshitser and B.R. Webber, Cavendish-HEP-95-2, Mar 1995,
     hep-ph/9504219.
\bi{AZ}
     R. Akhoury and V.I. Zakharov, Saclay-SPHT-95-043, Apr 1995,
     hep-ph/9504248.
\bi{mass}
     M. Beneke and V.M. Braun, Nucl. Phys. B426 (1994) 301;
\\   I. Bigi et al., Phys. Rev. D50 (1994) 2234.
\bi{GK}
     A.G. Grozin and G.P. Korchemsky, OUT-4102-53,
     Nov. 1994, hep-ph/9411323.
\bi{in}
     G.P. Korchemsky and G. Sterman, in preparation.
\eb

\end{document}